\documentclass[useAMS,usenatbib]{mn2e}

\usepackage{ulem} 
\usepackage{graphicx}
\usepackage{amsmath}
\usepackage{ulem}
\usepackage{rotating}
\usepackage{amssymb}
\usepackage[pdfborder={0 0 0},colorlinks=true,allcolors=blue]{hyperref}
\usepackage{aas_macros}

\def\lapprox{\hbox{\lower .8ex\hbox{$\,\buildrel < \over\sim\,$}}}
\def\gapprox{\hbox{\lower .8ex\hbox{$\,\buildrel > \over\sim\,$}}}

\title[The Y dwarf population with \textit{HST} -- I. First Results]{
The Y dwarf population with \textit{HST}: unlocking the secrets of our coolest neighbours -- I.
Overview \& First astrometric results\thanks{
Based on observations with the NASA/ESA {\it Hubble
Space Telescope}, obtained at the Space Telescope Science Institute,
which is operated by AURA, Inc., under NASA contract NAS 5-26555.
}
}
\author[C. Fontanive, L.\,R. Bedin \& D.\,C. Bardalez Gagliuffi]{
  C. Fontanive$^{1}$\thanks{E-mail: clemence.fontanive@csh.unibe.ch},
  L.\,R.\,Bedin$^{2}$\thanks{E-mail: luigi.bedin@oapd.inaf.it} and 
  D.\,C. Bardalez Gagliuffi$^{3}$  
\\  
$^{1}$Center for Space and Habitability, University of Bern, Gesellschaftsstrasse 6, 3012 Bern, Switzerland\\
$^{2}$INAF-Osservatorio Astronomico di Padova, Vicolo dell'Osservatorio 5, I-35122 Padova, Italy\\
$^{3}$American Museum of Natural History, 200 Central Park West, New York, NY 10024, USA 
}

\begin{document} 

\date{Accepted 2020 November 27. Received 2020 November 13; in original form 2020 October 5}

\pagerange{\pageref{firstpage}--\pageref{lastpage}} \pubyear{2020}

\maketitle
 
\label{firstpage}

\begin{abstract}
%
  In this paper we present our project that aims at determining accurate distances and proper motions for the Y brown dwarf population using the \textit{Hubble Space Telescope}. We validate the program with our first results, using a single new epoch of observations of the Y0pec dwarf WISE~J163940.83$-$684738.6.
  These new data allowed us to refine its proper motion and improve the accuracy of its parallax by a factor of three compared to previous determinations, now constrained to $\varpi=211.11\pm0.56$~mas. This newly derived absolute parallax corresponds to a distance of 4.737$\pm$0.013~pc, an exquisite and unprecedented precision for faint ultracool Y dwarfs.
\end{abstract}


%
\section{Introduction}
\label{introduction}
%
%
The new Y spectral sequence \citep{Cushing2011}, classifying objects with effective temperatures $<500$~K, is filling a crucial gap in mass and temperature between brown dwarfs and Jupiter, offering ideal proxies to study planet-like atmospheres \citep{Skemer2016}. However, fewer than $\sim$30 isolated Y dwarfs have been confirmed so far\footnote{\url{https://sites.google.com/view/ydwarfcompendium/home}}, and large disparities are seen within the uneven sets of limited data available, due to the challenges associated with astrometric and spectrophotometric observations for such intrinsically faint objects. The lack of reliable distance estimates to calibrate these systems is a key factor in our poor understanding of their physical and atmospheric properties. Indeed, accurate distances are vital for confident interpretations of measured quantities and assessments of inherent properties, which are thus indispensable to characterise these pivotal giant planet analogues. 

Dedicated programs aim at deriving trigonometric parallaxes for nearby brown dwarfs from the ground \citep{Tinney2014,Liu2016,Best2020} and from space \citep{Dupuy2013,Martin2018,Kirkpatrick2019}. However, the typical precision reached in these observationally-expensive campaigns (a few to tens of mas) results in substantial uncertainties in the underlying distances, and significant disagreements are observed between programs for the faintest targets (e.g. \citealp{Beichman2014}).

We present here the first results of our new \textit{Hubble Space Telescope} (\textit{HST}) project (GO~16229, PI: Fontanive), designed to measure the most precise parallaxes and proper motions to date for the 
majority of the ultracool Y dwarf population. \textit{HST} provides a unique platform to obtain precise astrometric parameters for Y dwarfs, with a remarkable astrometric precision and unmatched capabilities at near-infrared wavelengths for observations of such cold objects. By exploiting the \textit{Gaia} Data Release~2 (DR2; \citealp{GaiaCollaboration2016,GaiaCollaboration2018}) astrometric solutions for bright stars in \textit{HST} field of views, the exquisite astrometric precision of \textit{HST} can be translated into an absolute reference frame with an equally exquisite astrometric accuracy. In particular, the combination of \textit{HST} and \textit{Gaia}~DR2 in this way can enable the derivation of astrometric parameters with \textit{Gaia}-level precisions for sources like Y dwarfs, that are certainly too dim for \textit{Gaia}. In \citet{BedinFontanive2018,BedinFontanive2020}, we demonstrated that such an approach can achieve uncertainties of less than 2~mas on parallaxes and at the $\sim$0.3-mas level on proper motions, with a minimal three epochs of \textit{HST} observations acquired over $\sim$half a decade.

With this program, we will obtain the final epochs of observations required to implement this procedure for 19 Y dwarfs. Combined to archival data, these observations will allow us to derive reliable and uniform astrometric parameters for these objects, constraining distances and kinematics for the vast majority of the existing population of Y dwarfs.
In this paper, we use new \textit{HST} data of the Y0pec dwarf WISE~J163940.83$-$684738.6 to refine its parallax and proper motion and validate our approach. We introduce the target in Section \ref{target}, and describe the observations in Section~\ref{observations}. Section~\ref{analysis} presents our analysis, with an overview of our astrometric method from \citet{BedinFontanive2018,BedinFontanive2020} and the new results for the studied target. Our conclusions and outlook for the project are summarised in Section~\ref{conclusion}.

%
\section{W1639--6847} 
\label{target}
%

WISE~J163940.83$-$684738.6 (hereafter W1639$-$6847) is a nearby Y0pec ultracool brown dwarf, first discovered in \citet{Tinney2012} and subsequently classified by \citet{Tinney2014} and \citet{Schneider2015}. Model-derived physical parameters for this brown dwarf are highly uncertain and in some cases rather unrealistic \citep{Leggett2017,Zalesky2019}. This is likely due to a combination of the deviant spectrophometric features of this outlier brown dwarf, and the somewhat unreliable distance to the object available at the time of these studies (see \citealp{BedinFontanive2020} for details). 

We used this target in  \citet{BedinFontanive2020} to refine our \textit{HST}-based astrometric method (see Section~\ref{fitAP}) first presented in \citet{BedinFontanive2018}. W1639$-$6847 was a prime target for our approach, with three epochs of \textit{HST} observations taken between 2013 and 2019, and several independent -- but rather inconsistent -- measurements of its parallax and proper motion available in the literature \citep{Tinney2012,Tinney2014,Pinfield2014,Martin2018,Kirkpatrick2019}.

By anchoring these three epochs of \textit{HST} data acquired over $\sim$6~years to the \textit{Gaia}~DR2 reference frame, we were able to constrain its astrometric parameters to unprecedented \textit{Gaia}-like precisions \citep{BedinFontanive2020}. The achieved precisions of $\sim$2~mas on the parallax and at the sub-mas level on the proper motion already represented considerable improvements over previous estimates. Nonetheless, this parallax estimate for W1639$-$6847 relied entirely on the dataset with the lowest astrometric precision among the \textit{HST} epochs available at the time, which also did not provide an optimal coverage of the yearly parallax ellipse, leaving room for improvement on the astrometric parameters of this benchmark brown dwarf.

%
\section{Observations} 
\label{observations}
%

Previous \textit{HST} datasets for W1639$-$6847 consist of three separate epochs with a total of 18 infrared images, described in \citet{BedinFontanive2020}.
We recently acquired new \textit{HST} observations of this target as part of our dedicated program, GO~16229 (PI: Fontanive), with data taken on September 2 2020.

The new data consist of a single epoch made of 8 exposures in the F160W filter, collected over one full \textit{HST} orbit with the infrared (IR) channel of the Wide Field Camera~3 (WFC3) instrument. During the orbit, a large 8-point dither pattern was chosen, with a deep image obtained at each dithered position. The first four exposures were of duration 302.938~s each, followed by four exposures of 327.939~s. All images were acquired in MULTIACCUM mode with SAMP-SEQ=SPARS25, using NSAMP=13 samples for the former set of exposures and NSAMP=14 for the latter, for a total exposure time of 2523.508~s in the F160W band.
The observations were planned so as to keep the target at the center of the field of view (FoV) instead of remapping the FoV of previous datasets, a choice driven by the method described in \citet{BedinFontanive2018,BedinFontanive2020} which relies entirely on \textit{Gaia}~DR2 stars. 

A total of 26 individual images (18+8) are therefore employed for the analysis described in the following section.

%
\section{Analysis} 
\label{analysis}
%
The data reduction and analyses of this work follow the exact steps presented in \citet{BedinFontanive2018,BedinFontanive2020}, adding the data from the newest epoch of the science target, reduced in the same way as previous datasets.
We briefly summarise these methods below but refer the reader to those original works for detailed descriptions of the procedures and extensive discussions.

\subsection{Data Reduction}

Positions and magnitudes of detected sources were extracted in every WFC3/IR flat-fielded image from the newest \textit{HST} dataset, using the publicly available software developed by J. Anderson \citep{Anderson&King2006}. Measured positions in raw pixel coordinates were then corrected for the distortion of the camera\footnote{
Jay Anderson made publicly available the WFC3/IR distortion solution and the point-spread functions (PSFs) he derived in \citet{Anderson2016}, also providing softwares to use them. This material can be found at \url{https://www.stsci.edu/~jayander/WFC3/}, where the sub-directory \texttt{WFC3\_GC/} contains the distortion solution, and \texttt{WFC3\_PSFs} the PSFs for various filters.}.
For all sources sufficiently bright to be in the \textit{Gaia}~DR2 catalogue with full 5-parameter astrometric solutions, the positions of detected stars were then placed into the ICRS frame, transforming their \textit{Gaia} positions to the specific epoch of the \textit{HST} observations using the sources' parallaxes and proper motions, thus linking the \textit{HST} reference frame to the \textit{Gaia}~DR2 system. The procedure involves going back and forth to the tangential plane, using the methods and equations detailed in Section~3 of \citet{BedinFontanive2018}.

The same tasks were already performed for the previous observational datasets in \citet{BedinFontanive2020},  reregistering each frame onto the observational plane at epoch 2013.12, and the new data were similarly linked to the same reference system. This provides us with a common reference frame for all images from the various available epochs, defined by \textit{Gaia}~DR2 anchor sources that can be re-positioned to the relevant epochs from their respective \textit{Gaia}~DR2 astrometric solutions.
This allows us to transform the positions of every detected source to the absolute ICRS reference frame for every image and epoch, including sources much dimmer than those detectable by \textit{Gaia}, such as our extremely faint Y dwarf science target.

\subsection{Stack Images}

From these coordinate transformations into the common reference frame, we created stacked images for the latest observations of W1639$-$6847, complementing those from \citet{BedinFontanive2020} for the previous epochs and available online. We provide the newest stacks in the F160W filter as supplementary electronic material, saved in \texttt{fits} format, with the absolute astrometric solution provided in the header World Coordinate System keyword (see \citet{BedinFontanive2018,BedinFontanive2020} for additional details about the stacked images).

\subsection{Determination of the Astrometric Parameters}
\label{fitAP}

The 26 individual \textit{HST} images available for W1639$-$6847 each yield a 2D positional measurement of the brown dwarf. This provides us with a total of 52 individual data points to infer 5 astrometric parameters: the position ($\alpha$, $\delta$), proper motion ($\mu_{\alpha\cos\delta}$, $\mu_{\delta}$) and absolute parallax ($\varpi$). Following the methods described in \citet{BedinFontanive2018,BedinFontanive2020}, we used the Naval Observatory Vector Astrometry Software ({\tt NOVAS}) tool from the U.S. Naval Observatory \citep{Kaplan2011} to compute time-dependent positions for stars based on ICRS coordinates, proper motions and parallaxes. We then used a Levenberg-Marquardt algorithm \citep{More1980} to find the minimisation of the 5 parameters for W1639$-$6847.

\begin{figure*}
\begin{center}
\vspace{1cm}
\includegraphics[width=150mm]{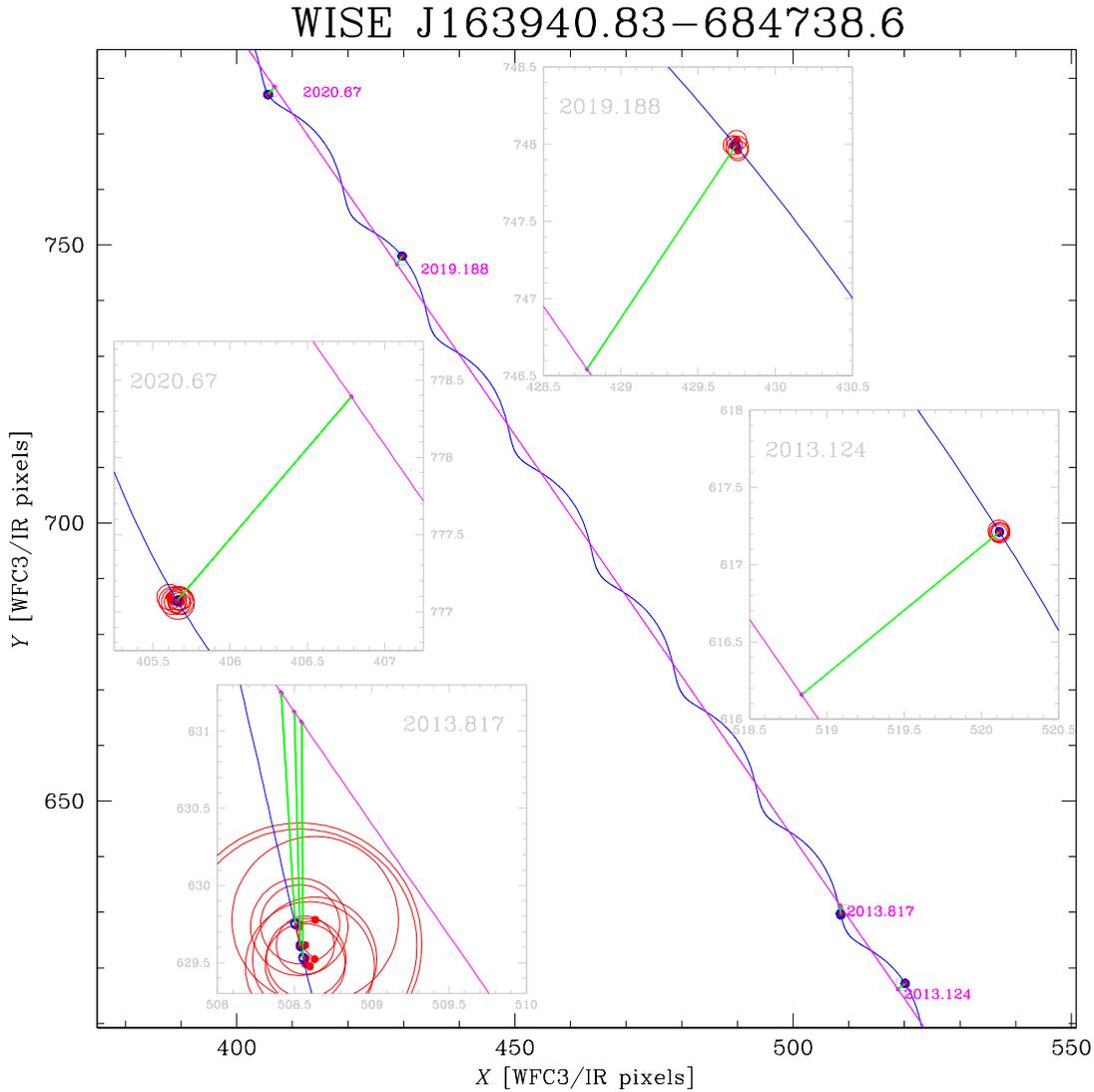}
\caption{
  Comparison of our astrometric solution (blue line) with the individual observed data points (red bullets) for W1639$-$6847 in the distortion-corrected observational plane at epoch 2013.12. The four major epochs are labeled, with insets labeled accordingly showing a more meaningful zoom-in of the data points. The sizes of the red circles indicate the quality-fit parameter \citep{Anderson2008} for each data point, with smaller radii for better measurements. To better highlight the parallax component of the motion, a magenta line marks the motion of an object with the same proper motion but placed at infinite distance (i.e. with zero parallax). Green lines show the parallax contributions at each epoch, and red segments connect the individual data points with their expected positions (blue bullets on the blue line) according to the best fit.
\label{cycloid}
}
\end{center}
\end{figure*}

\begin{figure*}
\begin{center}
\includegraphics[width=150mm]{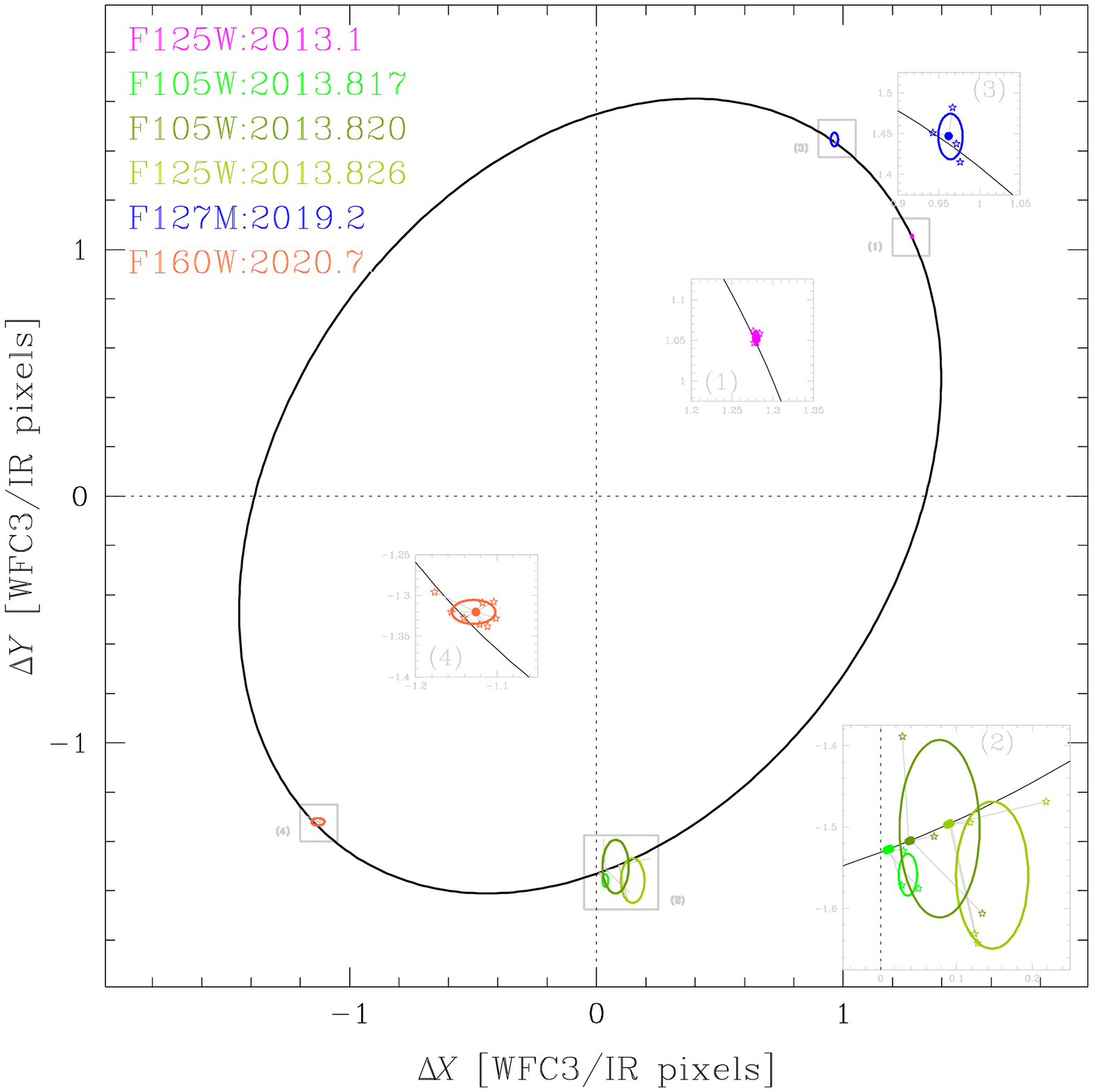}
\caption{
  Our solution for the parallax ellipse in the distortion-corrected reference coordinate system at epoch 2013.12. Individual \textit{HST} data points are indicated with star symbols, connected with small segments to their expected best-fit positions. Smaller ellipses indicate the 1-$\sigma$ spread of individual data points within each epoch. Note how ellipses are significantly smaller for the first and the last two epochs (magenta, blue and orange), compared to the 2013.8 sub-epochs (various shades of green). Insets in gray all have the same scale, and show zoom-in views around the locations marked by gray boxes.
\label{Ell}
}
\end{center}
\end{figure*}

Our results are presented in Table~\ref{tabASTR}, and compared in Table~\ref{lit} to our previous determinations and the \textit{Spitzer}-derived astrometry from \citet{Kirkpatrick2019} (see \citealp{BedinFontanive2020} for a complete compilation of astrometric measurements for W1639$-$6847 in the literature). The 0.05~mas added to the error budget of $\varpi$ is for the systematic uncertainties inherent to \textit{Gaia}~DR2 parallaxes \citep{Lindegren18}, although inconsequential compared to the estimated parallax error.
Figure~\ref{cycloid} shows our best-fit solution for the proper motion and parallax of the Y dwarf, along with the \textit{HST} measurements at each epoch. Figure~\ref{Ell} displays the parallax ellipse of the target, subtracting the proper motion from the total displacement. 

As discussed in \citet{BedinFontanive2020}, the parallax estimate derived in that work for W1639$-$6847 relied entirely on the epoch with the lowest astrometric precision (degraded by a nearby star and anomalously high background level; epoch plotted in green in Figure~\ref{Ell}), thus requiring an additional epoch of observations to be further validated and refined. We demonstrate here that our new epoch, optimally timed to measure the maximum parallax elongation (orange epoch in Figure~\ref{Ell}), is sufficient to improve the accuracy of the parallax by a factor of more than 3 over our previous estimate. As expected, improvements on proper motions and positions are, instead, marginal, since these quantities were already well constrained by two epochs widely-separated in time and with high astrometric accuracies taken at the same phase of the year (blue and magenta epochs).

We also tested the robustness of our astrometric solution by excluding the degraded epoch 2013.8 from our fit. The astrometric solution derived without this poorer dataset is fully consistent with our main results within uncertainties (parallax within 0.02~mas, and even less significant disparities for proper motions and positions). This is not surprising given that data points are weighted with the quality-of-fit parameter (see \citealp{BedinFontanive2018,BedinFontanive2020}), and the data points from the 2013.8 epoch thus have the lowest weight. With a lower significance towards our coverage of the full parallax elongation, now optimised by the first and last two observational epochs, this intermediate dataset has a negligible influence towards the final solution.

%
\begin{table}
\caption{Absolute astrometric parameters of W1639$-$6847 in the
      ICRS reference frame. Positions are given at 2000.0 and 2015.5.
      Note the larger errors for epoch 2000.0, as result of a large extrapolation to this epoch.}
\center
\begin{tabular}{lcc}
\hline \hline
Parameter & Value & Uncertainty \\
\hline
$\alpha_{2000.0}$ [ $^{\rm h}$ $^{\rm m}$ $^{\rm s}$ ]         &    16:39:39.730  & $\pm$ 12\,mas \\
$\delta_{2000.0}$ [ $^{\circ}$ $^{\prime}$ $^{\prime\prime}$ ] & $-$68:47:06.693  & $\pm$  4\,mas \\ [0.2cm]
$\alpha_{2000.0}$ [degrees] &  249.915540  & $\pm$ 12\,mas \\
$\delta_{2000.0}$ [degrees] &  -68.785192  & $\pm$  4\,mas \\ [0.2cm]
$\alpha_{2015.5}$ [degrees] &   249.9224084 & $\pm$ 4.5\,mas \\
$\delta_{2015.5}$ [degrees] & $-$68.79857156 & $\pm$ 0.8\,mas \\ [0.2cm]
$\mu_{\alpha\cos{\delta}}$ [mas yr$^{-1}$]  &  $+$576.94 & $\pm$ 0.22 \\
$\mu_{\delta}$ [mas yr$^{-1}$]              & $-$3108.48 & $\pm$ 0.21 \\ [0.2cm]
%
%
$\varpi$ [mas]              & 211.11     & $\pm$ 0.56 ($\pm$0.05)$^*$ \\
\hline
\multicolumn{3}{l}{$^*$ systematic uncertainties inherent to Gaia\,DR2 parallaxes}\\
\end{tabular}
\label{tabASTR}
\end{table} 
%

%
\section{Conclusions and Outlook}
\label{conclusion}
%

With a new epoch of \textit{HST} observations for W1639$-$6847 designed specifically to complement archival \textit{HST} data, we were able to constrain its astrometric parameters to unparalleled accuracies for such a faint ultracool brown dwarf. In particular, we achieved a 0.56-mas uncertainty on the parallax, a 3-fold improvement over our previous estimate without this newest epoch (\citealp{BedinFontanive2020}; see Table~\ref{lit}). This is also considerably better than the precisions of a few to tens of mas typically reached in distance measurements for nearby Y dwarfs \citep{Beichman2014,Martin2018,Kirkpatrick2019}. Such results are only possible thanks to the higher angular resolution of \textit{HST}/WFC3 compared to \textit{Spitzer}, and the approach of cross-registration with \textit{Gaia} sources. Our derived parallax corresponds to a remarkably precise distance of $4.737\pm0.013$~pc, which will incontestably be vital to investigate the peculiar nature and distinct atmospheric features of W1639$-$6847.

These results confirm the power of the method from \citet{BedinFontanive2018,BedinFontanive2020} to obtain highly-precise astrometric measurements for the very faintest astronomical objects, by linking \textit{HST} data to the \textit{Gaia} absolute reference frame, hence validating our program. This paper notably demonstrates the necessity of optimally-timed observations to achieve the best possible results on parallaxes and proper motions. Our ongoing \textit{HST} campaign will last about two years and will provide the new epochs needed to derive similar astrometric solutions for a total of 19 Y dwarfs. 
With observations carefully and strategically planned based on existing \textit{HST} epochs for each object, at times of year both maximising the coverage of the parallax ellipse and affording the best insights into proper motions, we anticipate comparable results for all targets in our sample.

Robust and uniform parallax measurements for the major part of the Y population will be essential to enhance our heavily scattered understanding of their characteristics, and improve theoretical models that severely lack empirical validation at the coldest temperatures \citep{Schneider2015}. Precise distances are indeed required for the knowledge of absolute fluxes, and therefore to estimate bolometric luminosities and unbiased spectral energy distributions. Measuring these quantities for benchmark objects to the highest levels of confidence will hence provide rich sources of testable fingerprints for theoretical models. The precise parallaxes anticipated from our program will also allow for detailed analyses of well-calibrated colour-magnitude diagrams, allowing for key probes of secondary effects like gravity or metallicity on the complex appearances of Y dwarfs, as well as the recognition of unresolved binarity.
Constraining the kinematics and fundamental properties of these objects will also be crucial in preparation of upcoming missions like \textit{JWST}, which, combined with robust calibrations, will offer invaluable new glimpses into these puzzling planetary atmospheres.

%
\begin{table*}
  \caption{
Comparison of our new absolute astrometric parameters for W1639$-$6847 with the most precise estimates in the literature.
    }
  \center
\begin{tabular}{lccccr}
\hline \hline
work  & $\mu_{\alpha^*}\pm\sigma\mu_{\alpha^*}$ & $\mu_\delta\pm\sigma\mu_\delta$ & $\varpi\pm\sigma_\varpi$ & $d$ & source\\
\#. authors (date) & [mas\,yr$^{-1}$] & [mas\,yr$^{-1}$] & [mas] & [pc] & facilities \\
\hline
   1. \citet{Kirkpatrick2019}   & $582.0\pm1.5$ & $-3099.8\pm1.5$ & $211.9\pm2.7$ & $4.72\pm0.06$ & \textit{Spitzer+Gaia\,DR1} \\
   2. \citet{BedinFontanive2020} & $577.21\pm0.24$ & $-3108.39\pm0.27$  & $210.35\pm1.82$ & $4.75\pm0.05$   & \textit{HST+Gaia\,DR2} \\
   3. \textbf{this work}        & $\mathbf{576.94\pm0.22}$ & $\mathbf{-3108.48\pm0.21}$  & $\mathbf{211.11\pm0.56}$ & $\mathbf{4.737\pm0.013}$   & \textit{\textbf{HST+Gaia\,DR2}} \\
\hline
\end{tabular}
\label{lit}
\end{table*} 
%

%
\section*{Acknowledgements}
We thank the anonymous referee for a thorough review of our work, a constructive discussion and useful comments.
CF acknowledges financial support from the Center for Space and Habitability (CSH).
LRB acknowledges support by MIUR under PRIN program \#2017Z2HSMF.
This work has been carried out within the framework of the NCCR PlanetS supported by the Swiss National Science Foundation.
This work is based on observations with the NASA/ESA Hubble Space
Telescope, obtained at the Space Telescope Science Institute, which is
operated by AURA, Inc., under NASA contract NAS 5-26555.
This work makes also use of results from the European Space Agency
(ESA) space mission Gaia. Gaia data are being processed by the Gaia
Data Processing and Analysis Consortium (DPAC). Funding for the DPAC
is provided by national institutions, in particular the institutions
participating in the Gaia MultiLateral Agreement (MLA). The Gaia
mission website is \url{https://www.cosmos.esa.int/gaia}. The Gaia
archive website is \url{https://archives.esac.esa.int/gaia}.
This research has benefitted from the Y Dwarf Compendium maintained by
Michael Cushing at \url{https://sites.google.com/view/ydwarfcompendium/}.
%

\section{Data Availability}

The reduced stacked images from this work are provided as supplementary electronic material.

\bibliographystyle{mnras}
\input{ms.bbl}

\label{lastpage}


\end{document}